\documentclass[preprint,showpacs,preprintnumbers,amsmath,amssymb, showkeys, pre]{revtex4}

\usepackage[dvips]{graphicx} 
\usepackage{epstopdf}
\usepackage{dcolumn} 
\usepackage{bm} 

\begin{document}

\preprint{APS/123-QED}

\title{Planewave diffraction at the periodically corrugated boundary \\
of vacuum and a negative phase--velocity material}

\author{Ricardo A. Depine}
\email{rdep@df.uba.ar}
\affiliation{Grupo de Electromagnetismo Aplicado\\  Departamento de F\'{\i}sica \\
Facultad de Ciencias Exactas y Naturales\\ Universidad de Buenos Aires \\
Ciudad Universitaria, Pabell\'{o}n I\\
1428 Buenos Aires, Argentina}

\author{Akhlesh Lakhtakia}
\email{AXL4@psu.edu}
\affiliation{Computational and Theoretical Materials Science Group \\
Department of Engineering Science and Mechanics \\
The Pennsylvania State University\\ University Park, PA 16802--6812, USA}


\begin{abstract}
Considering the diffraction of a plane wave by a periodically corrugated half--space,
we show that the transformation of the refracting medium
from positive/negative phase--velocity to negative/positive phase--velocity
type has an influence on the diffraction efficiencies. This effect increases
with increasing corrugation depth, owing to the presence of evanescent waves
in the troughs of the corrugated interface.

\end{abstract}



\maketitle

\newpage
\section{Introduction}
The rediscovery \cite{smith} of isotropic dielectric--magnetic materials exhibiting phase velocity vector opposed
in direction to
the time--averaged Poynting vector  has prompted a flurry of publications during the
last four years  \cite{LMW02,LMW03}. Many interesting effects have been predicted, with some
experimental backing as well \cite{GE02,PGLKT03,HBC03}.

Though several names have been proposed for this class of materials, we think that the
most descriptive is: { negative phase--velocity} (NPV) materials \cite{MLW}. In contrast, the
phase velocity and the time--averaged Poynting vectors are co--parallel in positive phase--velocity 
(PPV) materials. PPV materials are, of course, commonplace and require no introduction.

That the intrinsic difference between NPV and PPV materials has recognizable consequences
is easily gauged from a simple problem: reflection and refraction of a plane wave  due to
a homogeneously filled half--space. Let vacuum be the medium of incidence, while $\epsilon_2$
and $\mu_2$ denote the relative permittivity and relative permeability of
the medium of refraction. Let a linearly plane wave be incident on the planar interface of
the two mediums at an angle $\theta_0$, $(\vert\theta_0\vert<\pi/2)$, from the normal to the interface, and $\rho(\theta_0)$
be the reflection coefficient. If the transformation $\left\{\epsilon_{2}\to-\epsilon^\ast_{2},\,\mu_{2}\to-\mu^\ast_{2}\right\}$
is implemented, then $\rho(\theta_0)\to \rho^\ast(\theta_0)$, where the asterisk denotes the complex conjugate \cite{Lem}. Thus, the
replacement of a NPV/PPV half--space by an analogous PPV/NPV half--space changes the phase of the reflection
coefficient but not its magnitude. 

What would happen if the interface were to be corrugated \cite{donR}? Surface--relief gratings are periodically corrugated
surfaces that are
commonly used in electromagnetics, and many theoretical techniques are available to
compute their diffraction efficiencies \cite{May}.
Therefore, we decided to compute and compare the diffraction efficiencies of PPV and NPV surface--relief
gratings. In this report, we present our chief results here.
Section II contains a sketch of the theoretical method we chose, while
Section III is a discussion of the numerical results obtained. An $\exp(-i\omega t)$ time--dependence is implicit.

\section{Theory}
In a rectangular coordinate system $(x,y,z)$, we consider the periodically corrugated 
boundary $y=g(x)=g(x+d)$ between vacuum and a homogeneous, isotropic, linear
material, with $d$ being the corrugation period.
The region $y>g(x)$ is vacuous, whereas the medium occupying the region $y<g(x)$ is
characterized 
by complex--valued scalars 
$\epsilon_2=\epsilon_{2R} + i  \epsilon_{2I}$ and $\mu_2=\mu_{2R} + i\mu_{2I}$. 
If this medium is of the NPV type, then \cite{MLW, DL03}
\begin{eqnarray}
\epsilon_{2R} |\mu_2| + \mu_{2R} |\epsilon_2| <0  \,;  \label{conditrad2}
\end{eqnarray}
otherwise
\begin{eqnarray}
\epsilon_{2R} |\mu_2| + \mu_{2R} |\epsilon_2| >0  \,. 
\end{eqnarray}
A linearly polarized electromagnetic plane wave is incident on this boundary from the region 
$y>g(x)$ at an angle $\theta_0$, $(\vert\theta_0\vert<\pi/2)$, with respect to the $y$ axis. 

Let the function $f(x, y)$ represent the $z$--directed component of the total electric 
field for the $s$--polarization  case, and the $z$--directed component 
of the total magnetic field for the $p$--polarization case \cite{BW80}.
Outside the corrugations, $f(x, y)$ is rigorously represented by the following Rayleigh 
expansions \cite{donR}:
\begin{eqnarray}
f(x,y) &=& \exp\left[ i\,( \alpha_{0} x - \beta^{(1)}_0 y)\right] + \nonumber \\
&&\sum_{n=-\infty}^{+\infty} \rho_n \,\exp\left[i\,( \alpha_n x + \beta^{(1)}_{n} y)\right] \,,
\,\,\,\,\,\,\,\,\,\, y > \mbox{max}\,g(x) \,\,,\label{f1} 
\end{eqnarray}
\begin{eqnarray}
f(x,y) = 
\sum_{n=-\infty}^{+\infty} \tau_n \,\exp\left[i\,( \alpha_n x - \beta^{(2)}_{n} y)\right] \,,
\,\,\,\,\,\,\,\,\,\, y < \mbox{min}\,g(x) \,\,.  \label{f2} 
\end{eqnarray}
Here, $\left\{\rho_n\right\}_{n=-\infty}^{\;\;\;\;+ \infty}$ and 
$\left\{\tau_n\right\}_{n=-\infty}^{\;\;\;\;+ \infty}$ are scalar coefficients to be determined;  and
\begin{equation}
\left.
\begin{array}{ll}
\alpha_0=\frac{\omega}{c}\, \sin\theta_0\\[5pt]
\alpha_n = \alpha_0 + {2n\pi}/{d}\\
\beta^{(1)}_n = \sqrt{\frac{\omega^2}{c^2}  - \alpha_n^2} \\[5pt]
\beta^{(2)}_n = \sqrt{\frac{\omega^2}{c^2}\epsilon_2\,\mu_2 - \alpha_n^2}
\end{array}\right\}\,,
\end{equation}
where $c$ is the speed of light in vacuum. Note that $\beta^{(1)}_n$ is either purely real
or purely imaginary; and the conditions
\begin{equation}
\left.\begin{array}{ll}
{\rm Re} \left[\beta^{(1)}_n\right]  \geq 0\\[5pt]
{\rm Im} \left[\beta^{(1)}_n\right]  \geq 0
\end{array}\right\}\, \forall n
\end{equation}
are appropriate for plane waves in the vacuous half--space $y>g(x)$. The refracting half--space $y < g(x)$ being filled by
a material medium, $\epsilon_{2I}>0$ and $\mu_{2I} >0$ by virtue of causality. The refracted plane waves
must attenuate as $y\to-\infty$, which requirement leads to the condition
\begin{equation}
 {\rm Im}\left[\beta^{(2)}_n\right] > 0\, . \label{imbetan2}
 \end{equation}
Fulfilment of this condition automatically fixes the sign of ${\rm Re} \left[\beta^{(2)}_n\right] $, regardless of the signs of $\epsilon_{2R}$ and $\mu_{2R}$. 
We must note here that the transformation $\left\{ \epsilon_{2R}\to-\epsilon_{2R},
\mu_{2R}\to-\mu_{2R}\right\}$
alters the signs of the real parts of all $\beta_n^{(2)}$.

Boundary conditions at $y=g(x)$ require the continuity of the tangential components of the total
electric field   and the total magnetic field. Hence,
\begin{equation}
\left.\begin{array}{ll}
f(x,g(x)+) = f(x,g(x)-)\\
\hat{n}\cdot\nabla f(x,g(x)+)=\sigma^{-1}\,\hat{n}\cdot\nabla f(x,g(x)-)
\end{array}\right\}\,,
 \label{bcs} 
\end{equation}
where $\sigma=\mu_2$ for the $s$--polarization case and $\sigma=\epsilon_2$ for the $p$--polarization case, 
while $\hat{n}$ is a unit vector normal to the boundary.  

At this stage we invoke the Rayleigh hypothesis \cite{donR}~---~that is, we assume that expansions (\ref{f1})
and  (\ref{f2}), which are strictly valid outside the corrugated region, can be used in the boundary conditions (\ref{bcs}). 
Doing so, and then projecting into the Rayleigh basis $\left\{\exp(i \, \alpha_mx)\right\}_{m=-\infty}^{\;\;\;\;+ \infty}$, 
we obtain a system of linear equations for $\left\{\rho_n\right\}_{n=-\infty}^{\;\;\;\;+ \infty}$ and 
$\left\{\tau_n\right\}_{n=-\infty}^{\;\;\;\;+ \infty}$. Following Maradudin \cite[p. 427]{Mar}, we write
down the  system  in matrix form as 
\begin{eqnarray}
\left[ 
\begin{array}{cc}
{\cal M}_{11} & {\cal M}_{12} \cr
{\cal M}_{21} & {\cal M}_{22} \end{array}\right] \,
\left[ 
\begin{array}{c} 
{\cal R} \cr 
{\cal T}
\end{array}
\right]=
\left[
\begin{array}{c}
{\cal U} \cr
{\cal V}
\end{array}
\right] \,.\label{system}
\end{eqnarray}
The $(m,n)$--th elements of the four matrixes on the right side of (\ref{system}) are
\begin{equation}
\left.\begin{array}{ll}
{{\cal M}_{11}}{\Big\vert_{mn}}= - D_{mn}(\beta^{(1)}_n) \\[8pt]
{{\cal M}_{12}}{\Big\vert_{mn}}=   D_{mn}(-\beta^{(2)}_n) \\[8pt]
{{\cal M}_{21}}{\Big\vert_{mn}} =\beta^{(1)}_n\,D_{mn}(\beta^{(1)}_n) - \alpha_n\,E_{mn}(\beta^{(1)}_n)\\[8pt]
{{\cal M}_{22}}{\Big\vert_{mn}}= \frac{1}{\sigma}\left[
\beta^{(2)}_n\,D_{mn}(-\beta^{(2)}_n) + \alpha_n\,E_{mn}(-\beta^{(2)}_n)\right]
\end{array}\right\}\,,
\end{equation}
while the $m$--th elements of the  four column vectors in the same equation
are
\begin{equation}
\left.\begin{array}{ll}
{\cal R}{\Big\vert_{m}}=\rho_m\\[8pt]
{\cal T}{\Big\vert_{m}}=\tau_m\\[8pt]
{\cal U}{\Big\vert_{m}}=  D_{m0}(-\beta^{(1)}_0) \\[8pt]
{\cal V}{\Big\vert_{m}}=  \beta^{(1)}_0\,D_{m0}(-\beta^{(1)}_0) + \alpha_0\,E_{m0}(-\beta^{(1)}_0)
\end{array}\right\}\,.
\end{equation}
The integrals appearing in the foregoing equations are defined as
\begin{equation}
D_{mn}(u) = \frac{1}{d}\,\int_{0}^{d} \exp{[- i \, \frac{2\pi}{d} (m-n)\, x + iu
g(x) ]} \, dx \,\label{Dmn}
\end{equation}
and
\begin{equation}
E_{mn}(u) = \frac{1}{d}\,\int_{0}^{d} g^{\prime}(x)\exp{[- i \, \frac{2\pi}{d} (m-n)\, x + iu g(x) ]} \, dx \,, \label{Emn}
\end{equation}
with the prime denoting differentiation with respect to argument.

Equation (\ref{system}) has to be appropriately truncated 
and solved to determine the reflection coefficients $\rho_n$ and 
refraction coefficients $\tau_n$. Diffraction efficiencies 
\begin{eqnarray}
e_{n}^r = \frac{{\rm Re}\left[\beta^{(1)}_n\right]}{\beta^{(1)}_0}\,\vert\rho_n\vert^2 \,, \label{ern}
\end{eqnarray}
are defined
for the reflected orders. 
The normalized power absorbed across one period of the corrugated interface is given by
\begin{eqnarray}
P_a = 
{\rm Re} \Biggl[   
\frac{1}{\beta^{(1)}_0\,\sigma}\,\sum_{n,\;m} \Biggl \{
\alpha_n\,E_{mn} \Bigl[ \Bigl(\beta^{(2)}_m\Bigr)^\ast - \beta^{(2)}_n \Bigr]\,+ \nonumber\\
\beta^{(2)}_n \,D_{mn} \Bigl[\Bigl (\beta^{(2)}_m \Bigr)^\ast - \beta^{(2)}_n \Bigr]
\Biggl \} \,\tau_n\,\tau_m^\ast \Biggl]  \,. \label{prpa2}
\end{eqnarray}
The principle of conservation of energy requires that
\begin{eqnarray}
\sum_n e_{n}^r + P_a=1  \,. \label{prpa1}
\end{eqnarray}
When we implemented the procedure presented, we checked that the condition (\ref{prpa1}) 
was satisfied  to an error of 10~ppm. This was usually achieved by retaining 15 terms (i.e., $-7\le n \le 7$) 
in the Rayleigh expansions (\ref{f1}) and (\ref{f2}) of the fields. 

\section{Numerical Results and Discussion}
\begin{figure}[ht] 
\begin{center} 
\begin{tabular}{c}
\includegraphics[width=8.2cm]{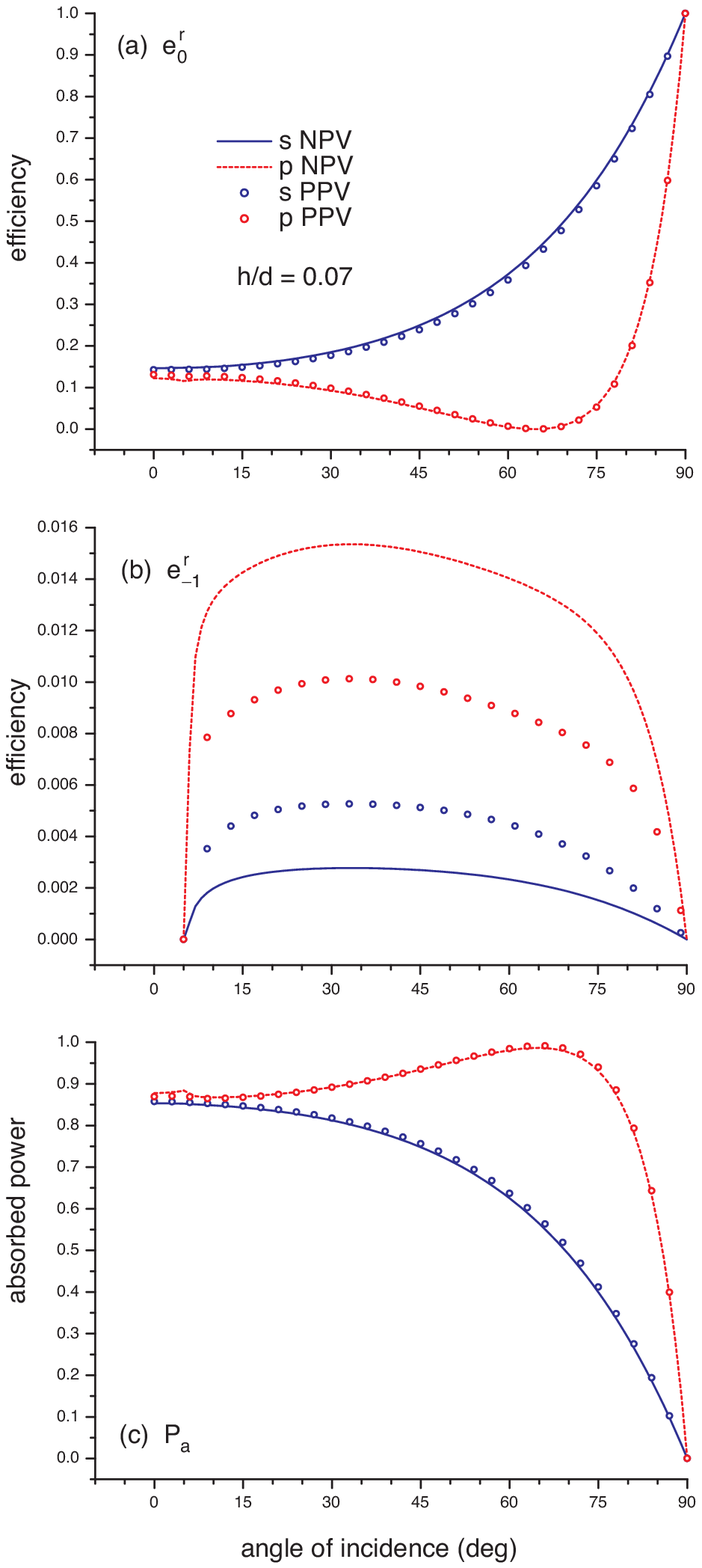} 
\end{tabular}
\end{center} 
\caption[example]{ \label{Fig1} Diffraction efficiencies $e_0^r$ and $e_{-1}^r$ as
well as the normalized absorbed power $P_a$ 
as functions of the incidence angle $\theta_0$, for a sinusoidally corrugated
interface between vacuum and a linear homogeneous medium. The interface function $g(x)=0.5\;h \cos(2\pi x/d)$,
where $h/d=0.07$ and $\omega d/c = 2\pi/1.1$. The refracting medium is of either the
PPV ($\epsilon_2=5+i0.01,\,\mu_2=1+i0.01$) or the NPV ($\epsilon_2=-5+i0.01,\,\mu_2=-1+i0.01$)
type. Calculations were made for both the $s$-- and the $p$--polarization cases.  }
 \end{figure}

\begin{figure}[ht] 
\begin{center} 
\begin{tabular}{c}
\includegraphics[width=8.2cm]{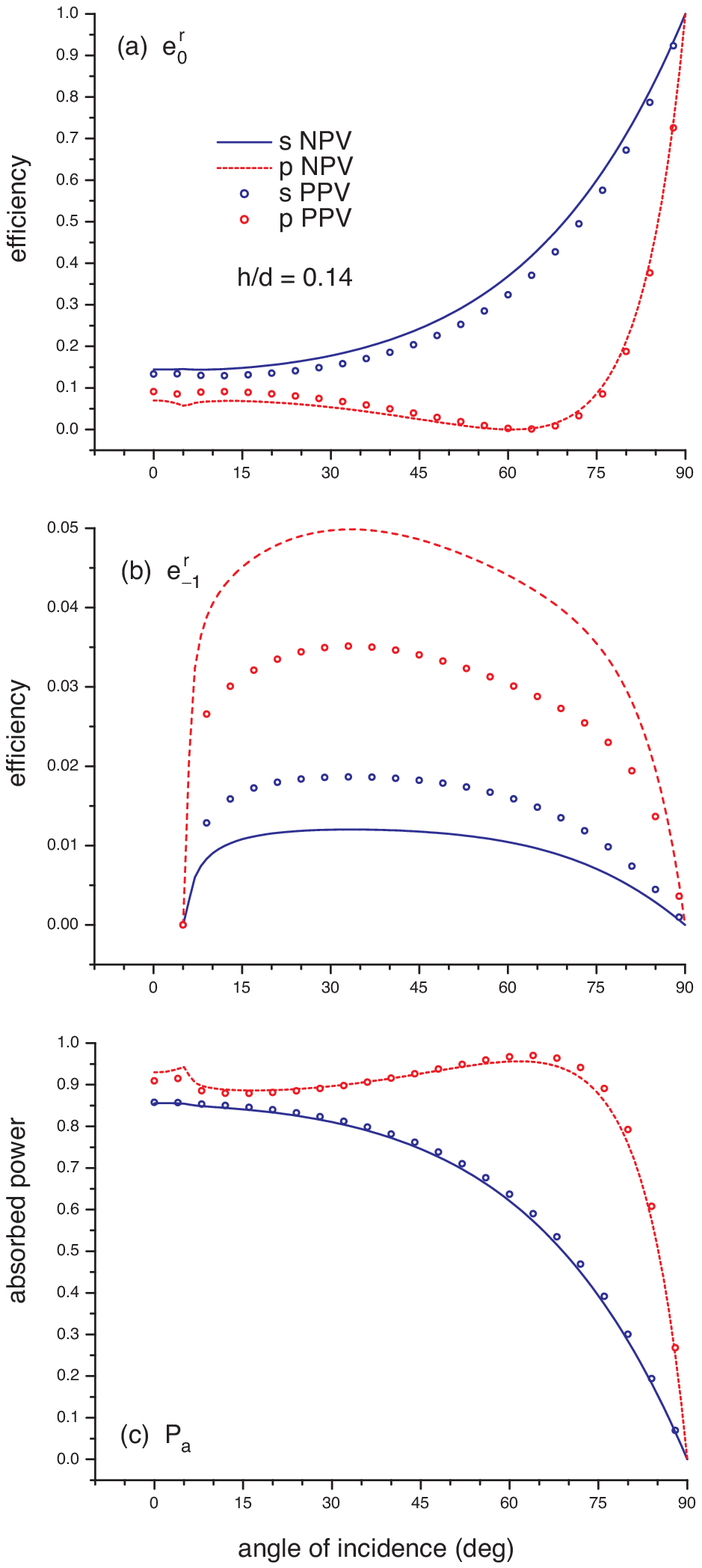} 
\end{tabular}
\end{center} 
\caption[example]{ \label{Fig2} Same as Figure \ref{Fig1}, but for $h/d=0.14$.}
 \end{figure}

\begin{figure}[ht] 
\begin{center} 
\begin{tabular}{c}
\includegraphics[width=8.2cm]{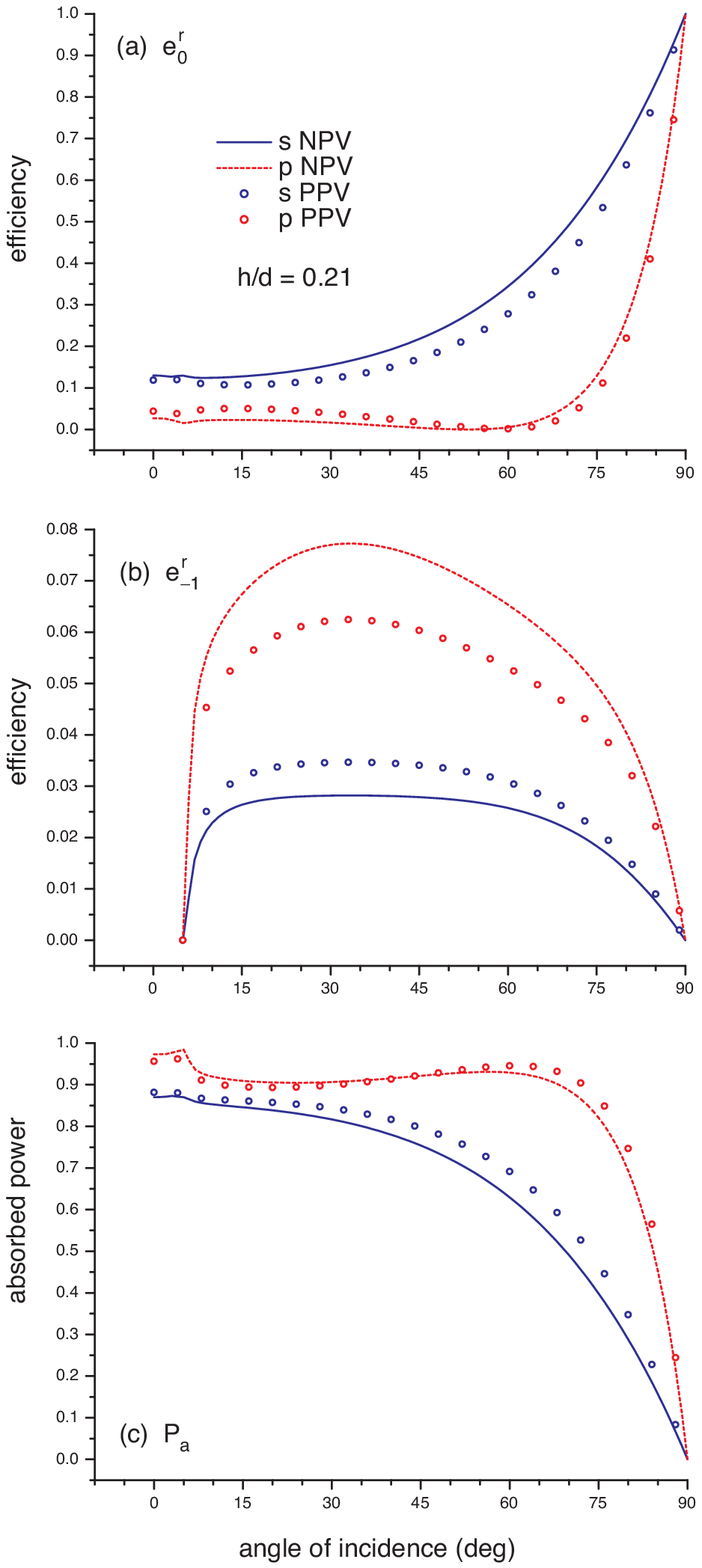} 
\end{tabular}
\end{center} 
\caption[example]{ \label{Fig3} Same as Figure \ref{Fig1}, but for $h/d=0.21$.}
 \end{figure}

We chose the corrugations to be sinusoidal: $g(x)=0.5\;h \cos(2\pi x/d)$. For this type of boundary between vacuum and a 
penetrable dielectric medium,
good results have been obtained for $h/d<0.3$ \cite{petit, hill}. 
We calculated diffraction efficiencies
for refracting mediums of both the PPV ($\epsilon_2=5+i0.01,\,\mu_2=1+i0.01$) and the NPV ($\epsilon_2=-5+i0.01,\,\mu_2=-1+i0.01$)
types. Calculations were made for both the $s$-- and the
$p$--polarization cases. Fixing the ratio $\omega d/c = 2\pi/1.1$, we
plotted the diffraction efficiencies $e_0^r$ and $e_{-1}^r$ 
as well as the absorption $P_a$ as functions of $\theta_0\in[0,\pi/2)$ for
$h/d=0.07$ (Figure \ref{Fig1}), $h/d=0.14$ (Figure \ref{Fig2}) and $h/d=0.21$ (Figure \ref{Fig3}). 

When $h/d=0$~---~i.e., when the interface is planar~---~it has been shown \cite{Lem}
that the transformation $\left\{ \epsilon_{2R}\to-\epsilon_{2R},
\mu_{2R}\to-\mu_{2R}\right\}$ does not change $e_0^r$. No wonder, the same transformation does not seem to be very effective
in affecting $e_0^r$ when $h/d=0.07$. As the corrugations grow deeper (i.e., as $h/d$ increases in value), the presented data shows
that the transformation of the refracting medium from NPV/PPV to PPV/NPV increasingly affects $e_{0}^r$ and $P_a$. 

Why should this be so? Now, for a planar interface, the transformation $\left\{ \epsilon_{2R}\to-\epsilon_{2R},
\mu_{2R}\to-\mu_{2R}\right\}$ leaves  
 the magnitude of the reflection coefficient {\em only} unchanged for non--evanescent
incident plane waves; but that is not a true statement for incident evanescent plane waves \cite{motl2}.
In the troughs of the corrugated interface, the total field that exists has both specular ($n=0$)
and nonspecular ($n\ne 0$) components.  Most of the nonspecular components are like evanescent 
plane waves because
they are characterized by ${\rm Re}\left[\beta_n^{(1)}\right]=0$. Their presence ensures that the diffraction efficiencies
are affected 
by the transformation of the refracting medium from NPV/PPV to PPV/NPV.

Before concluding, let us  point out that the numerical results presented here for   NPV surface--relief gratings 
agree with the results of a perturbational approach, thereby validating the limited use of the Rayleigh
hypothesis for NPV gratings in the same way as for PPV gratings \cite{DLpapnew}.
Also, the emergence of homogeneous NPV materials promises  new types of gratings which could be
significantly different from their PPV counterparts.

\begin{acknowledgments}
R.A.D. acknowledges financial support from Consejo Nacional de Investigaciones Cient\'{\i}ficas 
y T\'ecnicas (CONICET), Agencia Nacional de Promoci\'on Cient\'{\i}fica y Tecnol\'ogica (ANPCYT-BID 802/OC-AR03-04457) and 
Universidad de Buenos Aires (UBA). A.L. acknowledges partial support from the Penn State Materials   Research Science and Engineering Center.
\end{acknowledgments}

\newpage
 
\newpage

\newpage

\end{document}